\documentclass[twocolumn,superscriptaddress,amsmath,amssymb,aps]{revtex4-1}
\usepackage[dvipdfmx]{graphicx}
\usepackage[dvipdfmx]{color}
\usepackage{mathrsfs}
\usepackage{xr}
\usepackage{tabularx}
\usepackage{calculator}
\usepackage{numprint}

\newcommand{\ho}{\mbox{$\hat{O}$}}

\begin{document}
\title{Generalized Moment Method for Gap Estimation\\ and Quantum Monte Carlo Level Spectroscopy}
\author{Hidemaro Suwa}
\affiliation{Department of Physics, Boston University, 590 Commonwealth Avenue, Boston, Massachusetts 02215, USA}
\affiliation{Department of Physics, University of Tokyo, Tokyo 113-0033, Japan}
\author{Synge Todo}
\affiliation{Department of Physics, University of Tokyo, Tokyo 113-0033, Japan}
\affiliation{Institute for Solid State Physics, University of Tokyo, 
Kashiwa 277-8581, Japan}
\date{\today}
\begin{abstract}
We formulate a convergent sequence for the energy gap estimation in
the worldline quantum Monte Carlo method. The ambiguity left in the
conventional gap calculation for quantum systems is eliminated. Our
estimation will be unbiased in the low-temperature limit and also the
error bar is reliably estimated. The level spectroscopy
from quantum Monte Carlo data is developed as an application of the
unbiased gap estimation. From the spectral analysis, we precisely
determine the Kosterlitz-Thouless quantum phase-transition point of
the spin-Peierls model. It is established that the quantum phonon with a finite frequency is essential to the critical theory governed by the antiadiabatic limit, i.e.,\ the $k=1$ SU(2)
Wess-Zumino-Witten model.
\end{abstract}
\maketitle

The excitation gap is one of the most fundamental physical
quantities in quantum systems.  The Haldane phase and the $Z_2$
topological phase are characterized by the topologically protected
gap\,\cite{HasanK2010}. Recently the existence of gapful/gapless
quantum spin-liquid phases has been discussed in frustrated spin
systems\,\cite{Balents2010}. Not only in the gapful but also critical
phases, the system-size dependence of the excitation gap is useful for
the analysis of the quantum phase transition\,\cite{Nomura1995}. 
Particularly, the energy gap $\Delta$ in the conformal quantum phases
scales as $\Delta \propto x v / L$, apart from possible logarithmic
correction, where $L$ is the system size, $x$ is the scaling
dimension, and $v$ is the velocity appearing in the conformal field
theory\,\cite{Cardy1996,*FrancescoMS1997}.  An unbiased gap calculation thus allows for
extracting the universal properties of the critical phases from
finite-size data.

The gap estimation for large systems is not trivial.  For small
systems, it is possible to calculate the gap by the exact
diagonalization method. The reachable system size is, however,
strongly limited because of the explosion of required memory size and
computation time. The density matrix renormalization group (DMRG)
method\,\cite{White1992} works well for many one-dimensional systems,
but it becomes less effective in gapless or degenerated phases. In the
meanwhile, the quantum Monte Carlo (QMC) method based on the worldline
representation is a powerful method for various strongly correlated
systems without dimensional
restriction\,\cite{KawashimaH2004,*Sandvik2010}.
In previous QMC calculations\,\cite{Yamamoto1995,*MengLWAM2010}, the gap is extracted by
the fitting of the correlation function; the tail of the exponential
function is estimated as a fitting parameter
[see Eq.\,(\ref{eqn:c-tau}) below]. Here we encounter a trade-off
between the systematic error and the statistical error. The lower the
temperature is in a QMC simulation, the smaller the systematic error
becomes but the larger the statistical error does. It is because the
correlation in long imaginary time has an exponentially small absolute
value. Since in practice we do not have a prior knowledge for the
optimal temperature and the range of imaginary time where the
correlation function follows the asymptotic form, a choice of data
necessarily introduces some bias in the fitting procedure.
Our purpose in the present paper is to establish a versatile and
unbiased gap-estimation procedure free from an ambiguous fitting.

In the meanwhile, the recently advanced technology has allowed for
quantum simulators that can realize ideal quantum many-body
systems\,\cite{BlattR2012,*GeorgescuAN2014}. In particular, the quantum
phonon effect of trapped ions has caught a great deal of attention,
which provides rich physics and engineering, e.g., spin
frustration\,\cite{BermudezASRP2011}, long-range spin
interaction\,\cite{BrittonSKWFUBB2012}, phonon
superfluids\,\cite{PorrasC2004}, quantum gates\,\cite{ZhuMD2006},
etc. As an application of the present gap-estimation method, we will
elucidate the quantum phase transition of the spin system coupled
with quantum phonons, which is called the spin-Peierls
transition\,\cite{CrossF1979,KubokiF1987,CaronM1996,WelleinFK1998,WeisseWF1999,WeisseHBF2006,CitroOG2005}
and is accessible in the ion system\,\cite{BermudezP2012}. The level
spectroscopy\,\cite{Nomura1995} from Monte Carlo data is developed to
overcome the difficulty of the [Kosterlitz-Thouless (KT)] transition
that makes the conventional approaches ineffective. We establish that
the quantum phonon effect is essential to the spin-Peierls system and
its critical phenomena.

The spectral information of a quantum system described by a Hamiltonian $H$ is encoded in the
imaginary time (dynamical) correlation function:
\begin{eqnarray}
C(\tau) &=& \langle \hat{O}(\tau) \hat{O}^{\dagger} \rangle = \frac{1}{Z}{\rm tr} \left[ e^{\tau H} \hat{O} e^{-\tau H} \hat{O}^{\dagger} e^{- \beta H} \right] \nonumber \\
&=& \frac{1}{Z} \sum_{\ell, \ell'} b_{\ell, \ell'} e^{-\tau ( E_{\ell} - E_{\ell'} )} e^{-\beta E_{\ell'}} \nonumber \\
&\rightarrow& \sum_{ \ell \geq 1 } b_{\ell} e^{-\tau ( E_{\ell} - E_0 )} \quad ( \beta \rightarrow \infty ) \label{eqn:c-tau} ,
\end{eqnarray}
where $\ho$ is a chosen operator, $Z$ is the partition function and
$\beta = 1/T$ is the inverse temperature.  Also, $\{ | \ell \rangle
\}$ is the complete orthogonal set of eigenstates, $ E_\ell $ is the
associated eigenenergy, $b_{\ell,\ell'}=| \langle \ell' | \hat{O} |
\ell \rangle |^2$, and $b_{\ell} = b_{\ell,0}$,
where the ground state is $| 0 \rangle $, the first
excited state is $| 1 \rangle $, and so are the higher excited states, respectively.
We assume $ \ho | 0 \rangle \neq 0$, $ \langle 0 | \ho | 0
\rangle = 0 $, and $E_{\ell} > E_0$ ($\ell \geq 1$), i.e., the ground
state is certainly excited by the operator and the gap is finite.
The latter is the case for finite-size systems even if the ground state is degenerated in the thermodynamic limit.

Here, let us consider the moment of the imaginary time correlation
function\,\cite{Todo2006}:
\begin{eqnarray}
I_k = \frac{1}{k!} \int_0^{\infty} \tau^{k}C(\tau)  d\tau = \sum_{\ell \geq 1} \frac{ b_{\ell} }{ \Delta_{\ell}^{k+1} } \qquad ( k \geq 0 ) \label{eqn:I_k},
\end{eqnarray}
where $\Delta_{\ell} \equiv \Delta_{\ell,0}$, and
$\Delta_{\ell, \ell'} = E_{\ell} - E_{\ell'} $.  The higher
moment will be dominated by the contribution from the first excitation
gap as $I_k \sim b_1 / \Delta_1^{k+1}$ because $\Delta_1 < \Delta_2 <
\Delta_3 < \cdots$.  Then we can see a useful limit, $ ( I_{k} /
I_{k+m} )^{1/m} \rightarrow \Delta_1$ ($k \rightarrow
\infty$) $\forall m \in {\mathbf N}$.  However, we cannot use the
moment directly in finite-temperature simulations. It is because the
correlation function is periodic for bosons or anti-periodic for
fermions and the moment is not well defined. Then the Fourier
series can be exploited instead. Let us think of the bosonic
case because we will investigate spin excitation. The Fourier component of the correlation function at a Matsubara frequency $\omega_j =
2 \pi j / \beta$ $(j \in {\mathbf Z})$ is expressed as
\begin{eqnarray}
&&  \tilde{C}(\omega_j) = \int_0^{\beta} C(\tau) e^{i \tau \omega_j } d\tau \label{eqn:c-omega} \\
&&= \left\{
\begin{array}{l}
\displaystyle \frac{1}{Z} \sum_{\ell, \ell' } \frac{ g_{\ell, \ell'} 
  \Delta_{\ell, \ell'} }{ \Delta_{\ell, \ell'}^2 + \omega_j^2 } \quad (
\omega_j \neq 0 ) \\ 
\displaystyle \frac{1}{Z} \{ \sum_{E_{\ell} \neq
  E_{\ell'}} \frac{ g_{\ell, \ell'} }{ \Delta_{\ell, \ell'} } + \sum_{E_{\ell} =
  E_{\ell'}} b_{\ell, \ell'} e^{- \beta E_{\ell}} \beta  \} \quad (
\omega_j = 0 ) \nonumber ,
\end{array}
\right.
\end{eqnarray}
where $g_{\ell, \ell'} = b_{\ell, \ell'} ( e^{-\beta E_{\ell'}} -
e^{-\beta E_{\ell} } )$.  In many simulations, the so-called {\em
  second moment}\,\cite{CooperFP1982} is used as the lowest-order gap estimator:
\begin{eqnarray}
  \hskip-1em
  \hat{\Delta}_{(1, \beta)} = \omega_1 \sqrt{ \frac{ \tilde{C}(\omega_1)}{\tilde{C}(\omega_0) - \tilde{C}(\omega_1)}} \ \rightarrow \ \sqrt{\frac{I_0}{I_2}} \quad (\beta \rightarrow \infty) \label{eqn:delta-1} .
\end{eqnarray}
Interestingly, this estimator will be the ratio of the zeroth and the
second moment in the low-temperature limit.  We have to
take notice of the systematic error carefully. The error remains even
in $\beta \rightarrow \infty$ as
\begin{eqnarray}
\frac{ \hat{\Delta}_{(1, \beta)} }{ \Delta_1} \rightarrow \ 1 + \frac{1}{2} \sum_{\ell > 1} \left[ \frac{ b_{\ell} }{b_1} \frac{\Delta_{1}}{\Delta_{\ell}} + O\left( \left( \frac{\Delta_{1}}{\Delta_{\ell}} \right)^2 \right) \right] \label{eqn:delta-1-e},
\end{eqnarray}
which is typically a few percent of $\Delta_1$\,\cite{TodoK2001}.
This correction hampers proper identification of the universality
class in the level spectroscopy analysis as we will see below.

Our main idea in the present paper is to construct a sequence of gap
estimators that converges to a ratio of higher-order moments in the
low-temperature limit. We will consider an estimator that has a correction of $O ( ( \Delta_{1} / \Delta_{\ell} )^{2n-1} )$. Let us expand Eq.\,(\ref{eqn:c-omega})
in powers of ($1/\beta \Delta_{\ell,\ell'}$) and make a linear combination of Fourier components so that the lowest orders cancel;
$(-1)^n \sum_{k=0}^n x_{n,k} \tilde{C}(\omega_k) = \sum_{\ell, \ell'} 
g_{\ell,\ell'} \omega_1^{2n} \Delta_{\ell,\ell'}^{-(2n+1)} / Z + O( \beta^{-(2n+2)}
\Delta_{\ell,\ell'}^{-(2n+3)})$ with coefficients $x_{n,k}$. It
will be dominated by the smallest gap $\Delta_1$ in $\beta \rightarrow
\infty$ and $n \rightarrow \infty$. The coefficients $x_{n,k}$ satisfy the
following equations: $\sum_{k=0}^{n} x_{n,k} k^{2m} = \delta_{m,n}$ $(0
\leq m \leq n)$, where $\delta_{mn}$ is the Kronecker delta.
They are exactly solved for by the formula of the inverse of Vandermonde's
matrix\,\cite{Neagoe1996}. We can also show $(-1)^{n-1} \sum_{k=0}^n k^2 x_{n,k}
\tilde{C}(\omega_k) = \sum_{\ell, \ell'} g_{\ell,\ell'}
\omega_1^{2(n-1)} \Delta_{\ell,\ell'}^{-(2n-1)} / Z +
O( \beta^{-2n} \Delta_{\ell,\ell'}^{-(2n+1)})$ by using the same $x_{n,k}$. Then, a
sequence of the higher-order estimators is derived as
\begin{eqnarray}
  \hskip-2em
&\hat{\Delta}_{(n, \beta)}& = \omega_1 \sqrt{ - \left.  \sum_{k=0}^{n} k^2 x_{n,k} \tilde{C}( \omega_k ) \right/ \sum_{k=0}^{n} x_{n,k} \tilde{C}( \omega_k ) } \label{eqn:delta-n}
\end{eqnarray}
with $x_{n,k} = 1 / \prod_{j=0, j \neq k}^n ( k + j ) ( k - j )$.
Importantly,
$\hat{\Delta}_{(n,\beta)} \rightarrow \sqrt{ I_{2(n-1)} / I_{2n}} \quad (\beta \rightarrow \infty)$,
and the systematic error is expressed as
\begin{eqnarray}
  \hskip-1em
\frac{ \hat{\Delta}_{(n,\beta)}}{\Delta_1} \rightarrow  1 + \frac{1}{2} \sum_{\ell > 1} \left[ \frac{ b_{\ell} }{b_1} \left( \frac{\Delta_{1}}{\Delta_{\ell}} \right)^{2n-1} \! \! \! \! \! \! \! \! \! + O\left( \left( \frac{\Delta_{1}}{\Delta_{\ell}} \right)^{2n} \right) \right] .
  \end{eqnarray}
Note that $I_k$ can be achieved only for even-numbered $k$ since $\tilde{C}(\omega_j)$ is real.  As examples, $(x_{2,0}, x_{2,1}, x_{2,2}) = ( \frac{1}{4}, -\frac{1}{3},
\frac{1}{12} )$ for $n=2$\,\cite{TodoK2001}, and $(x_{3,0}, x_{3,1}, x_{3,2}, x_{3,3})
= ( - \frac{1}{36}, \frac{1}{24}, - \frac{1}{60}, \frac{1}{360} )$ for
$n=3$.  We have analytically written down the bias of the gap estimator\,(\ref{eqn:delta-n}) and shown the following remarkable property:
\begin{eqnarray}
\lim_{n\rightarrow \infty} \lim_{\beta \rightarrow \infty} \hat{\Delta}_{(n, \beta)} = \lim_{\beta \rightarrow \infty} \lim_{n \rightarrow \infty} \hat{\Delta}_{(n, \beta)} = \Delta_1 \label{eqn:lim-lim}.
\end{eqnarray}
That is, these two limits are interchangeable (see the Supplemental Material\,\footnote{See Supplemental Material attached below for the
  derivation of the systematic error in the case of discrete/continuum
  spectrum, the error-bar comparison to the fitting approach, and the
  recipe of the error optimization.} for details). This important property
makes our gap estimation greatly robust.   Note that the present approach works also in the stochastic series expansion QMC method by
the time generation\,\cite{SandvikSC1997}.

Our generalized gap estimator is applicable to any quantum
system.  As an example, we will show an application with the
level spectroscopy to the following one-dimensional $S=1/2$
spin-Peierls model:
\begin{eqnarray}
\hspace{-4mm}
H = \sum_r ( 1 + \sqrt{\frac{ \omega\, \lambda }{2} }( a_r + a_r^{\dagger} ) ) S_{r+1}
\cdot S_r + \sum_r \omega a_r^{\dagger}a_r,
\label{eqn:sp-model}
\end{eqnarray}
where $\omega$ is a dispersionless phonon frequency, $\lambda$ is a
spin-phonon coupling constant, $S_r$ is the spin-$\frac{1}{2}$
operator, $a_r$ and $a_r^{\dagger}$ are the annihilation and creation
operator of the soft-core bosons (phonons) at site $r$, respectively.  This
spin-Peierls model has been investigated in the adiabatic limit
($\omega \rightarrow 0$)\,\cite{CrossF1979}, from the antiadiabatic
limit ($\omega \rightarrow
\infty$)\,\cite{KubokiF1987,WelleinFK1998,WeisseWF1999,
  CaronM1996,WeisseHBF2006,PearsonBB2010,SandvikC1999}, and in its
crossover\,\cite{CitroOG2005}.  The relevance of the present model to
real materials, such as CuGeO$_3$\,\cite{HaseTU1993}, has been
discussed\,\cite{UhrigS1996}.
The model is expected to exhibit a KT-type
quantum phase transition between the Tomonaga-Luttinger (TL) liquid phase and the dimer
phase at a finite spin-phonon coupling\,\cite{CaronM1996,CitroOG2005},
which is absent when either spin or phonon is classically
treated\,\cite{CrossF1979}.  The realization of the quantum phase
transition was recently proposed in the trapped ion
system\,\cite{BermudezP2012}.

It is difficult to precisely locate the transition point by the
conventional analyses. The huge Hilbert space with the soft-core
bosons hinders the sufficient-size calculation by the diagonalization
method\,\cite{WelleinFK1998}. The effective spin-model approach by the
perturbation\,\cite{KubokiF1987} or the unitary
transformation\,\cite{WeisseWF1999} does not take into account all
marginal terms, e.g., the 4-spin and 6-spin interactions examined in
Ref.\,\cite{TangS2011}.  As for the DMRG method, it needs an additional symmetry
breaking term, which blurs the phase transition point, in the degenerated phase\,\cite{PearsonBB2010}.
In addition, the method has difficulty in precise calculation of the relevant quantities, such as the central charge, around an essential singularity\,\cite{ChenXMCCY2013}.
Also, the previous QMC approach\,\cite{SandvikC1999} suffers from the exponential divergence of the correlation length and the logarithmic correction around the KT transition point. These
difficulties mentioned above can be overcome by employing the level
spectroscopy method\,\cite{Nomura1995,Tzeng2012} combined with our precise gap
estimation.
In the TL liquid phase, the both of triplet and singlet
excitation are gapless in the thermodynamic limit, but the lowest
excited state of finite-size systems is the
triplet because of the logarithmic correction\,\cite{AffleckGSZ1989,*SinghFS1989,*GiamarchiS1989,*Eggert1996}. In the dimer phase, on the other hand, the
first excited state is the singlet for finite-size systems.  It forms
the degenerated ground states eventually in the thermodynamic
limit. Thus the excitation gaps of the triplet and singlet excitation
intersect at a spin-phonon coupling for finite-size systems.
The transition point can be efficiently extrapolated from the gap-crossing points\,\cite{Nomura1995}.

\begin{figure}
\begin{center}
  \includegraphics[width=8cm]{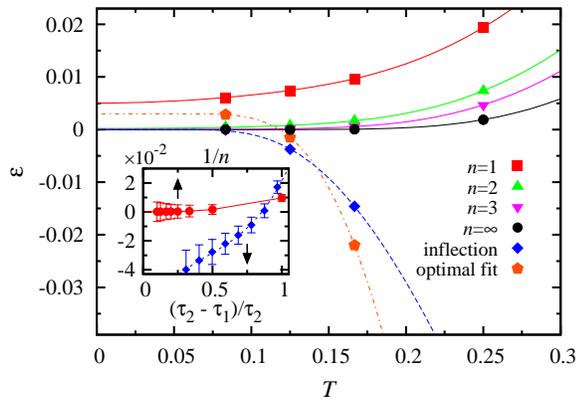}
\caption{Triplet-gap estimation (relative) error of our
  estimators\,(\ref{eqn:delta-n}) ($n=1,2,3, \infty$), the
  inflection-point value of $ - d \log C(\tau) / d \tau $, and the
  optimal fit (defined in the main text) for the spin-Peierls
  model\,(\ref{eqn:sp-model}) with $L=4$, $\omega=4$, $\lambda=1/2$,
  $D=4$, which has $\Delta_1 \simeq 1.111388$.
  The inset shows the $n$ dependence of the gap estimate
  (circles) for $1 \leq n \leq 10 $ and the $\tau_1$ dependence of the
  $\log C(\tau)$ linear-fit result (diamonds) for $\tau_1 \leq \tau
  \leq \tau_2$ with $\tau_2=\beta/4$, calculated from $2^{20}(\sim 10^6)$ Monte Carlo steps at $\beta=6$.
}
\label{fig:tg}
\end{center}
\end{figure}
We used the continuous-time worldline representation\,\cite{BeardW1996,SandvikSC1997}
and the worm (directed-loop)
algorithm\,\cite{ProkovievST1998,*SyljuasenS2002} in the QMC
method. Thanks to the exponential form of the diagonal operators, our simulation is free from an
occupation-number cutoff of the soft-core bosons. The Fourier
components of the correlation function\,(\ref{eqn:c-omega}) are directly
calculated during the simulation. The worm-scattering probability is
optimized in rejection (bounce) rate by breaking the detailed
balance\,\cite{SuwaT2010}.  The boundary condition was periodic in the
space and time directions. More than $2^{25}$($\simeq 3.4 \times 10^7$) Monte Carlo samples were taken in total after $2^{18}$($\simeq 2.6 \times 10^5$) thermalization steps.  The error bar of the gap estimates is calculated by the jackknife analysis\,\cite{Berg2004}.

First, the convergence of our gap estimate was tested for $L=4$,
$\omega=4$, $\lambda=1/2$, where $L$ is the system size. We set
$\omega$ here is fairly larger than the actual spin gap because this
condition is satisfied for large systems in the relevant spin-phonon
coupling region.  The boson occupation number cutoff $D$ was set to 4
only in this test for comparing with the diagonalization result.
Fig.\,\ref{fig:tg} shows the calculated triplet-gap estimation errors,
where $\ho = \sum_r S^z_r e^{i\pi r}$ is used in the dynamical
correlation function. We compared the gap
estimators\,(\ref{eqn:delta-n}) to the previous
approach\,\cite{Yamamoto1995,*MengLWAM2010} where the first gap is
estimated as $- d\log C(\tau) / d \tau $ from the asymptotic
form\,(\ref{eqn:c-tau}). The derivative will show a plateau at the gap
value in an appropriate $\tau$ region. When $\beta(=1/T)$ is not large
enough, however, the plateau is indistinct. Then the inflection point
could be used, but it is hard to estimate in practice (here we
calculated it by longer QMC simulation for comparison). As an
another practical and reasonable gap estimation, we test a linear fit
of $\log C(\tau)$ for $\tau_1 \leq \tau \leq \tau_2$, where we fix
$\tau_2=\beta/4$. The inset of Fig.\,\ref{fig:tg} shows the feasible
convergence of the gap estimate in $n$ and the difficulty of finding
appropriate $\tau_1$ for the linear fit. The function $\log C(\tau)$
is poorly fitted to a linear form at small $\tau_1$, while it has larger
statistical error at large $\tau_1$. Then the gap error resulting from
the linear regression takes a minimum value at optimal $\tau^*_1$, which
we call ``optimal fit.'' Even though it seems reasonable, the optimal fit
underestimates the gap at $\beta=6,8$ and overestimates it at
$\beta=12$ as shown in the main panel of the figure. Meanwhile,
the second-moment estimator\,($n=1$) has a non-negligible bias even in $T \rightarrow 0$ as expected from
Eq.\,(\ref{eqn:delta-1-e}).  The estimate with large enough $n$ (we call it the $n=\infty$ estimate hereafter), on
the other hand, exponentially converges to the exact value as the
temperature decreases\,\cite{Note1}.  The bias convergence is much faster
than that of the inflection point (one of the best estimates from
the fitting approach). Moreover, the higher-order estimator provides a reliable error bar, while the optimal fit significantly underestimates it\,\cite{Note1}.
Therefore, our approach is more precise and straightforward than the fitting approach.
In the present study, we have used a simple
recipe to optimize $n$ and $\beta$, minimizing both the systematic and
the statistical error\,\cite{Note1}.

\begin{figure}[t]
\begin{center}
  \includegraphics[width=7cm]{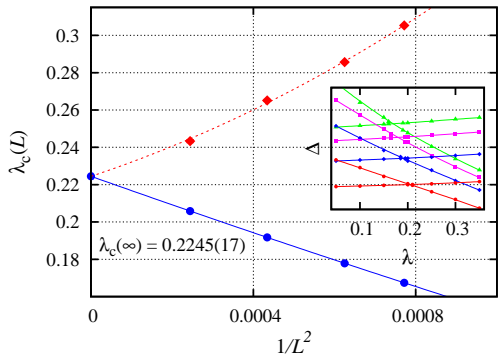}
  \caption{Convergence of the gap-crossing point (circles) between the
  triplet and the singlet excitation for $L=36,40,48,64$, together
  with the crossing point of the spin susceptibility (diamonds)
  between $\chi_s(L)/L$ and $\chi_s(L/2)/(L/2)$. The spin-phonon
  coupling dependence of the gaps is shown in the inset for each $L$.
  The dashed line is the fitting curve
  with $\lambda_{\rm c}(\infty)$ fixed, which results in
  large $\chi^2 /$dof $\approx 5.0$. The statistical errors are smaller than
  the symbol size.}
\label{fig:tg-sg}
\end{center}
\end{figure}
The scaling of the gap-crossing point for the spin-Peierls model between the triplet and singlet
excitation is shown in Fig.\,\ref{fig:tg-sg}.
For the singlet excitation gap, we
used $\ho = \sum_r S_r \cdot S_{r+1} e^{i\pi r}$.  The bare excitation
phonon gap was set to $\omega=1/4$ for the comparison with the
previous result\,\cite{SandvikC1999}. The transition point $\lambda_{\rm c} = 0.2245(17)$ in the
thermodynamic limit was extrapolated
without logarithmic correction, which is much
more precise than the previous estimate, $0.176 < \lambda_{\rm c}
< 0.23 $\,\cite{SandvikC1999} in our notation. Also the spin
susceptibility $\chi_s=\int_0^{\beta} \sum_{r} \langle S^z_r (\tau)
S^z_{0} \rangle e^{i\pi r}d\tau $ could be used for finding the
transition point (Fig.\,\ref{fig:tg-sg}).
Nevertheless, the gap-crossing point provides the much more reliable extrapolation with $1/L^2$ correction from irrelevant fields\,\cite{Nomura1995}, while the susceptibility is likely to have some more complicated corrections.

\begin{figure}[t]
\begin{center}
  \includegraphics[width=7.5cm]{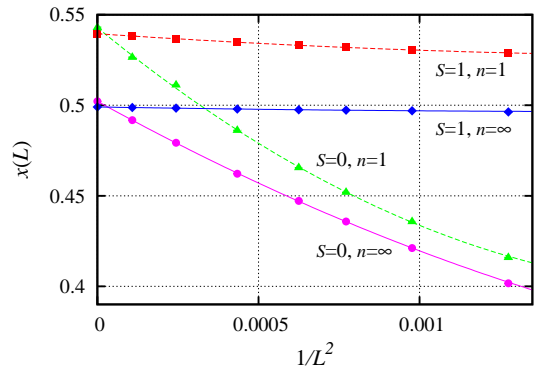}
\caption{System-size dependence of the scaling dimension corresponding
  to the triplet or the singlet excitation at the transition point
  ($\lambda=0.2245$), calculated from the second-moment ($n=1$) or the
  $n=\infty$ gap estimate.
}
\label{fig:dim}
\end{center}
\end{figure}

We have also calculated the velocity, the central charge, and the
scaling dimensions at the transition point, fixing
$\lambda=0.2245$. The velocity $v=1.485(8)$ was calculated from
the scaling form $v(L) = \Delta_{k_1} / k_1 = v + a / L^{2} + b / L^4
+ o(1/L^{4})$, where $\Delta_{k_1}$ is the triplet gap at
$k_1=2\pi/L$, $a$ and $b$ are non-universal constants.  The central
charge $c=0.987(13)$ was obtained from the finite-size
correction\,\cite{AffleckGSZ1989}, $E_0(L) = E_0 - \pi v c / 6L + o(1/L)
$.  The scaling
dimension corresponding to the triplet or singlet excitation was
calculated from the relation $x(L) = L \Delta_\pi / 2 \pi v$, where
$\Delta_\pi$ is the lowest (triplet or singlet) excitation gap at
$k=\pi$.  As shown in Fig.\,\ref{fig:dim}, the $n=\infty$ estimates converged to $x_{S=1}=0.502(3)$ and $x_{S=0}=0.499(3)$ without logarithmic correction as expected only at the transition point\,\cite{Nomura1995}. Hence we conclude that
this transition point is described by the $k=1$ SU(2)
Wess-Zumino-Witten model\,\cite{Witten1984} with $c=1$ and
$x=1/2$. On the other hand, the second-moment estimates ($n=1$)
failed to approach 1/2 as seen in Fig.\,\ref{fig:dim}. This identification of the
critical theory clearly demonstrates the importance of the
higher-order estimator. 
The present study non-trivially clarified that the critical theory at
the transition point of the spin-Peierls model with a finite phonon frequency coincides with that in the antiadiabatic limit ($\omega \rightarrow \infty$) where the effective spin model is the frustrated $J_1$-$J_2$ chain\,\cite{Nomura1995}.  Our result strongly indicates that the quantum phonon effect is {\em relevant} to the spin-Peierls system in the sense that it necessarily triggers the universal KT phase transition.

In conclusion, we have presented the generalized moment method for the gap estimation.
The advantages of our method over the previous approaches are as follows: the unbiased estimation [Eq.\,(\ref{eqn:lim-lim})], the absence of ambiguous procedure, the
faster convergence w.r.t.\ the temperature, and the reliable error-bar
estimation. We emphasize that our approach is
generally applicable to any quantum system. The QMC level spectroscopy
was demonstrated, for the first time, for the KT transition in the
spin-Peierls model. This spectral analysis will likely work in various systems including most conformal phases. We elucidated that the quantum phonon
effect is relevant to the critical theory of the spin-phonon
system, which is expected to be universal in many kinds of one-dimensional systems, e.g., (spinless) fermion-phonon systems, by virtue of the well-established transformations.
The clarified quantum phase transition and the criticality would be directly observed in the quantum simulator\,\cite{BermudezP2012}.

The authors are grateful to Anders W. Sandvik, Thomas C. Lang, and Hiroshi Ueda for the valuable discussion. The most simulations in the present paper were done by using the facility of the Supercomputer Center, Institute for Solid State Physics, University of Tokyo. We also used the computational resources at the Institute for Information Management and Communication, Kyoto University and Computing and Communications Center, Kyushu University through the HPCI System Research Project (No.\,hp140204, hp140162). The simulation code has been developed based on the ALPS library\,\cite{ALPS2011s}. The authors acknowledge the support by KAKENHI (No.\,23540438, 26400384) from JSPS and CMSI in SPIRE from MEXT, Japan. HS is supported by the JSPS Postdoctoral Fellowships for Research Abroad.

\clearpage
\widetext
\renewcommand{\theequation}{S\arabic{equation}}
\renewcommand{\thefigure}{S\arabic{figure}}
\renewcommand{\citenumfont}[1]{S#1}
\renewcommand{\bibnumfmt}[1]{[S#1]}
\setcounter{equation}{0}
\setcounter{figure}{0}

\begin{center}
  \textbf{
    \begin{large}
  Supplemental Material: Generalized Moment Method for Gap Estimation\\
  and Quantum Monte Carlo Level Spectroscopy
    \end{large}
  }
\end{center}
\section{Systematic error of higher-order gap estimate}
We provide here the detailed form of the bias of the gap estimator~(\ref{eqn:delta-n}) and its limiting form in $n \rightarrow \infty$ and $\beta \rightarrow \infty$. The gap estimator is rewritten as
\begin{align}
\hat{\Delta}_{(n,\beta)} = \sqrt{ \frac{ \displaystyle \sum_{ E_{\ell} \neq E_{\ell'}} g_{\ell, \ell'} \Delta_{\ell,\ell'}  \prod_{j=1}^n \frac{1}{ \Delta^2_{\ell,\ell'} + \omega_j^2 } } { \displaystyle \sum_{ E_{\ell} \neq E_{\ell'}} \frac{ g_{\ell, \ell'} }{ \Delta_{\ell,\ell'} } \prod_{j=1}^n \frac{1}{ \Delta_{\ell,\ell'}^2 + \omega_j^2 } + \sum_{E_{\ell}=E_{\ell'}} b_{\ell,\ell'} \beta e^{-\beta E_{\ell}} \prod_{j=1}^n \frac{1}{ \omega_j^2} }} .
\end{align}
In the equation, as the
reminder of the definitions, $\Delta_{\ell, \ell'} = E_{\ell} - E_{\ell'} $,
$b_{\ell, \ell'} = | \langle \ell' | \hat{O} | \ell \rangle |^2$,
$g_{\ell,\ell'}= b_{\ell, \ell'} ( e^{-\beta E_{\ell'}} - e^{-\beta E_{\ell} } )$,
$\omega_j = 2 \pi j / \beta$ $(j \in {\mathbf
  Z})$.
Here we used the useful equations:
\begin{align}
  \sum_{k=0}^n \frac{ x_{n,k} }{ \Delta^2 + \omega^2_k } &= (-1)^n \omega^{2n}_1 \prod_{k=0}^n \frac{ 1 }{ \Delta^2 + \omega^2_k } \label{eqn:sum-prod} , \\
  \sum_{k=0}^n \frac{ k^2 x_{n,k} }{ \Delta^2 + \omega^2_k } &= (-1)^{n-1} \omega^{2(n-1)}_1 \prod_{k=1}^n \frac{ 1 }{ \Delta^2 + \omega^2_k } \label{eqn:sum-prod2} ,
\end{align}
where $x_{n,k} = 1 / \prod_{j=0, j \neq k}^n ( k + j ) ( k - j ) $
as defined in the main text. These equations are derived from the imposed condition
on $x_{n,k}$ to cancel the lowest orders of $(1/ \beta \Delta_{\ell,\ell'})$ as explained in the main text:
\begin{align}
  \sum_{k=0}^{n} x_{n,k} k^{2m} = \delta_{m,n} \quad (0 \leq m \leq n)
\quad  &\Longleftrightarrow \quad
\sum_{k=0}^n x_{n,k} \sigma_{n,m} (-k) = (-1)^n \delta_{m,n} \quad (0 \leq m \leq n) \label{eqn:esp} \\
&\, \Longrightarrow \quad
  \sum_{k=1}^n k^2 x_{n,k} \sigma_{n,m} (-k) = (-1)^{n-1} \delta_{m,n-1} \quad (0 \leq m \leq n-1) \label{eqn:esp2} ,
\end{align}
where $\sigma_{n,m} (-k) = \sigma_m ( 0, 1, 2^2, 3^2, ..., (k-1)^2, (k+1)^2, ..., n^2)$, and
$\sigma_m$ is the elementary symmetric polynomial of order $m$, e.g., $\sigma_0 (a_1, a_2, a_3)=1$, $\sigma_1 (a_1, a_2, a_3) = a_1 +
a_2 + a_3$, $ \sigma_2 (a_1, a_2, a_3) = a_1 a_2 + a_2 a_3 + a_3 a_1$,
$\sigma_3 ( a_1, a_2, a_3 ) = a_1 a_2 a_3$.

The systematic error of the higher-order gap estimate is expressed as
\begin{equation}
  \frac{ \hat{\Delta}_{(n, \beta)} }{ \Delta_1} = \sqrt{ \frac{ 1 + R_n (\beta) } {  1 + \displaystyle  F_n (\beta)  + D_n (\beta) } }
\quad  \rightarrow  \quad \ 1  + \frac{1}{2} \sum_{\ell > 1} \left[ \frac{ b_{\ell} }{b_1} \left( \frac{\Delta_{1}}{\Delta_{\ell}} \right)^{2n-1} \! \! \! \! \! \! + O\left( \left( \frac{\Delta_{1}}{\Delta_{\ell}} \right)^{2n} \right) \right] \quad ( \beta \rightarrow \infty ) \label{eqn:delta-n-e} ,
\end{equation}
where
\begin{align}
R_n (\beta) &= \sum_{ E_{\ell} \neq E_{\ell'}, ( \ell, \ell' ) \neq ( 1, 0 ) } \left( \frac{g_{\ell, \ell'}}{g_{1,0}} \right) \left( \frac{ \Delta_{1} }{ \Delta_{\ell,\ell'} } \right)^{-1} h(n, \beta, \Delta_{\ell, \ell'} ) , \\
F_n (\beta) &= \sum_{ E_{\ell} \neq E_{\ell'}, ( \ell, \ell' ) \neq ( 1, 0 ) } \left( \frac{g_{\ell, \ell'}}{g_{1,0}} \right) \left( \frac{ \Delta_{1} }{ \Delta_{\ell,\ell'} } \right) h(n, \beta, \Delta_{\ell, \ell'} ) , \\
D_n (\beta) &=  \sum_{E_{\ell} = E_{\ell'} }  \left( \frac{b_{\ell, \ell'}}{g_{1,0}} \right) \beta \Delta_1 \, e^{ - \beta E_{\ell} } \prod_{j=1}^{n} \left( 1 + \frac{ \Delta_1^2 }{ \omega_j^2 } \right) , \\
&h(n,\beta,\Delta_{\ell, \ell'}) = \prod_{j=1}^n \frac{ \Delta^2_1 + \omega^2_j }{ \Delta^2_{\ell, \ell'} + \omega^2_j  } , 
\end{align}
the summations for $F_n$ and $R_n$ are taken over $E_{\ell} \neq
E_{\ell'}$ except $(\ell, \ell') = ( 1, 0 )$ and $ ( 0, 1 )$, $D_n$
term comes from $x_{n,0} \, \tilde{C}(0)$, and $b_{\ell} = b_{\ell,0}$.
We showed, in the main text,
the limiting form: $\lim_{\beta \rightarrow \infty} \hat{\Delta}_{(n,
  \beta)} = \sqrt{ I_{2(n-1)} / I_{2n} }$, where $I_k$ is the moment defined as
Eq.~(\ref{eqn:I_k}).  Then, $ \lim_{n\rightarrow \infty} \lim_{\beta
  \rightarrow \infty} \hat{\Delta}_{(n, \beta)} = \Delta_1$ since $ (
I_{k} / I_{k+m} )^{1/m} \rightarrow \Delta_1 $ ($k
\rightarrow \infty$) $\forall m \in {\mathbf N}$.

Next, let us
consider the limiting form in $n \rightarrow \infty$ at a finite
temperature. We use the product expansion form of the hyperbolic
function: $\sinh ( \pi z ) = \pi z \prod_{j=1}^{\infty} ( 1 + z^2 /
j^2 )$ with $z=\beta \Delta_{\ell,\ell'}/2\pi$ in Eq.~(\ref{eqn:delta-n-e}). The finite-$n$ corrections are expressed as
\begin{equation}
  \begin{split}
 \prod_{j=1}^{n} ( 1
+ z^2 / j^2 ) &\sim \frac{ \sinh ( \pi z ) }{ \pi z } \exp \left\{ - z^2 \left( \frac{1}{n+1} + \frac{1}{2(n+1)(n+2) } \right) \right\} \\
&\sim \frac{ \sinh (\pi z) }{ \pi z } \left[ 1 - z^2 \left( \frac{ 1 }{n+1} + \frac{1}{2 (n+1) (n+2) } \right) + \frac{ z^4}{2 (n+1)^2 } \right] \label{eqn:finite-n} .
\end{split}
\end{equation}
Here the asymptotic expansion of the Riemann zeta function was used;
\begin{align}
\sum_{k=1}^n \frac{ 1 }{ k^2 } = \frac{ \pi^2 }{ 6 } - \frac{ 1 }{ n + 1 } - \frac{ 1 }{ 2 (n+1) (n+2) } - \cdots .
\end{align}
The limiting form is then expressed as
\begin{align}
\frac{ \hat{\Delta}_{(n, \beta)} }{ \Delta_1} = \sqrt{ \frac{ 1 + \displaystyle  R_{\infty} (\beta) } {  1 + G_{\infty} (\beta) } } + O \left( \frac{1}{n} \right) \label{eqn:delta-inf} ,
\end{align}
where
\begin{align}
R_{\infty} (\beta) &= \sum_{ E_{\ell} \neq E_{\ell'}, ( \ell, \ell' ) \neq ( 1, 0 ) }
\left( \frac{b_{\ell, \ell'}}{b_1} \right) \left( \frac{ \Delta_{1} }{ \Delta_{\ell,\ell'}  } \right)^{-2} e^{ - \frac{\beta}{2}{ \Delta_{\rm ex} } } , \\
G_{\infty} (\beta) &= \sum_{ (\ell, \ell') \neq ( 1, 0 ) } \left( \frac{b_{\ell, \ell'}}{b_1} \right) e^{ - \frac{\beta}{2}{ \Delta_{\rm ex} } } ,
\end{align}
and $\Delta_{\rm ex} = E_{\ell} + E_{\ell'} - E_1 - E_0 $.  Note that
the correction terms coming from $F_n$ and $D_n$ are included in
$G_{\infty}$. Then, $\lim_{\beta \rightarrow \infty} \lim_{n
  \rightarrow \infty} \hat{\Delta}_{(n, \beta)} = \Delta_1$ because
$\Delta_{\rm ex} > 0$. As a result, the finite-$n$ corrections are $O(1/n)$. It is attainable to confirm the convergence as shown in the inset of 
Fig.~\ref{fig:tg} in the main text. Furthermore, the $n=\infty$ estimate shows the exponential
convergence w.r.t. the temperature, which was indeed observed
for the test case in the main text (see Fig.~\ref{fig:tg}).

\section{Asymptotic Behavior of Gap Estimate for Continuum Spectrum}
The systematic error~(\ref{eqn:delta-n-e}) was formulated for a
discrete spectrum. This is the case in finite-size quantum systems
with a reasonable basis. We will show a generalization to a continuum
spectrum that will be achieved in the thermodynamic limit. In
practice, the crossover from the discrete to continuum case will be
observed in fairly large systems.  First, let us generalize the
formulation of the correlation functions. They are expressed as
\begin{align}
  C(\tau) &= \int_{-\infty}^{\infty} d\/ \epsilon S(\epsilon) e^{- \tau \epsilon} , \\
  \tilde{C}(\omega_j) &= \int_0^{\beta} d\/ \tau C(\tau) e^{ i \tau \omega_j} = \frac{1}{2\pi} \int_{-\infty}^{\infty} d\/ \epsilon \frac{ A(\epsilon) }{ \epsilon - i \omega_j } ,
\end{align}
where 
$A(\epsilon) = 2 \pi S(\epsilon) ( 1 - \zeta e^{ - \beta \epsilon} ) $ is the spectral function with $\zeta = 1$ (bosons or spins) or $\zeta=-1$ (fermions). 
The formulation of a discrete spectrum is recovered by setting
\begin{align}
  S(\epsilon) = \frac{1}{Z} \sum_{\ell,\ell'} b_{\ell,\ell'} e^{ - \beta E_{\ell'} } \delta( \epsilon - ( E_{\ell} - E_{\ell'}) ) .
\end{align}
Then the gap estimator~(\ref{eqn:delta-n}) is generally expressed as
\begin{align}
\hat{\Delta}_{(n,\beta)} = \sqrt{ \frac{ \displaystyle \int_{- \infty}^{\infty} d \epsilon \, A(\epsilon) \epsilon \prod_{j=1}^n \frac{1}{ \epsilon^2 + \omega_j^2 } }{ \displaystyle \int_{- \infty}^{\infty} d \epsilon \, \frac{ A(\epsilon) }{ \epsilon } \prod_{j=1}^n \frac{1}{ \epsilon^2 + \omega_j^2 } } } \label{eqn:delta-conti}.
\end{align}
Let us focus on the spin case here.  The integrand with $O=S_i^z$ or
$S_i^z S_j^z$ ($i$ and $j$ are site indices) in Eq.~(\ref{eqn:delta-conti}) is symmetric at $\epsilon=0$, so considering only for $\epsilon \geq 0 $ suffices. Let us then write $A(\epsilon)=A_+(\epsilon)$ ($\epsilon \geq 0$).
When there is a delta peak at the first gap ($b_1>0$) and a finite gap between the first gap ($\Delta_1$) and the
second gap ($\Delta_2$), as $\Delta_2 - \Delta_1 > 0$, the systematic
error~(\ref{eqn:delta-n-e}) is expressed in a similar way converting
the discrete summation to the corresponding integral. The asymptotic form
becomes
\begin{align}
  \frac{ \hat{\Delta}_{(\infty,\beta)} }{ \Delta_1 } = 1 + O\left( e^{- \frac{\beta}{2} ( \Delta_2 - \Delta_1 ) }\right) , \\
  \frac{ \hat{\Delta}_{(n,\infty)} }{ \Delta_1 } = 1 + O\left( \left( \frac{\Delta_1}{\Delta_2}\right)^{2n-1} \right) .
\end{align}

Next let us think of the case where the system has a continuum
spectrum above the lowest gap ($\Delta_1$). The
asymptotic behaviors of the gap estimate
for some typical spectral function (with $\Delta_1>0$ or $\Delta_1=0$) are
summed up in Table~\ref{tab:asympt-conti}. The parameter constraint
for each case comes from the bounded correlation function:
$C(\tau) < \infty$.
  
  \vskip1.2em
  \underline{(i) $ A_+ (\epsilon) \sim ( \epsilon - \Delta_1 )^a \theta( \epsilon - \Delta_1 ) \quad (a>-1)$}

  \vskip0.3em
  When the spectrum is in a power of $\epsilon - \Delta_1$, the systematic error will be asymptotically $O(1/n)$ or $O(1/\beta)$; that is,
  \begin{align}
            \hat{\Delta}_{(\infty,\beta)} &\sim \sqrt{\frac{ Q(a,2)}{ Q(a,0) } } = \Delta_1 \left[ 1 + O \left( \frac{ 1 }{ \beta } \right) \right] , \\
    \hat{\Delta}_{(n,\infty)} &\sim \sqrt{\frac{ T(a,2n-1)}{ T(a,2n+1) } } = \Delta_1 \left[ 1 + O \left( \frac{ 1 }{ n } \right) \right] ,
  \end{align}
  where
  \begin{align}
    Q(a,m) & \equiv \int_{\Delta_1}^{\epsilon_M} d \epsilon \, \frac{ ( \epsilon - \Delta_1 )^a \epsilon^m }{ \sinh \left( \frac{ \beta \epsilon }{ 2 } \right) } \sim \int_{\Delta_1}^{\epsilon_M} d \epsilon \, 2 ( \epsilon - \Delta_1 )^a \epsilon^m e^{-\frac{ \beta \epsilon }{ 2 } } = 2^{a+2} \frac{ \Delta_1^m}{ \beta^{a+1} } e^{ - \frac{ \beta \Delta_1 }{ 2 } } \int_0^{q_M} dq \, q^a \left( 1 + \frac{2q}{\beta \Delta_1} \right)^m e^{-q} \nonumber \\
    & = \frac{ \Delta_1^m }{ \beta^{a+1} } e^{ - \frac{ \beta \Delta_1 }{ 2 } } \left( Q_0 + \frac{ Q_1 }{ \beta } \right) \quad ( \beta \gg 1 / \Delta_1 ),  \label{eqn:Q} \\
    T(a,m) & \equiv \int_{\Delta_1}^{\epsilon_M} d \epsilon \, \frac{ ( \epsilon - \Delta_1 )^a }{ \epsilon^m} = \int_0^{T_M} d t \, \frac{ \Delta_1^{a+1-m}}{m^{a+1}} \frac{ t^a}{ ( 1 + \frac{t}{m})^m} \sim \int_0^{T_M} d t \, \frac{ \Delta_1^{a+1-m}}{m^{a+1}} \frac{ t^a}{ e^{t} ( 1 - \frac{ t^2 }{ 2m } ) } \nonumber \\
    & = \frac{ \Delta_1^{a+1-m}}{m^{a+1}} \left( T_0 + \frac{ T_1 }{ m } \right) \quad ( m \gg 1 ) \label{eqn:T} .
  \end{align}
  In Eq.~(\ref{eqn:Q}) $q=\beta ( \epsilon - \Delta_1 ) / 2$, in Eq.~(\ref{eqn:T}) $\epsilon = (1+\frac{t}{m})\Delta_1$, and $\epsilon_M, q_M, t_M, Q_0, Q_1, T_0, T_1$ are constants.

    \begin{table}[t]
    \parbox{0.95\textwidth}{\caption{Asymptotic behavior of the gap estimate for typical continuum spectrum. In the spectral function row, $\theta(x)$ is the unit step function; it takes $0 \ (x < 0)$ or $1 \ ( x \geq 0)$. \label{tab:asympt-conti}}}
    {\renewcommand{\arraystretch}{2}
    \begin{tabularx}{0.95\textwidth}{@{} *5{>{\centering\arraybackslash}X}@{}}
      \hline \hline
      $A_+(\epsilon)$ & $( \epsilon - \Delta_1 )^a \theta( \epsilon - \Delta_1 )$ ($a>-1$) & $e^{ - c / (\epsilon - \Delta_1)^a} \theta( \epsilon - \Delta_1 )$ ($a,c>0$) & $  \epsilon^a$ ($a>0$) & $ e^{ - c / \epsilon^a} $ ($a,c>0$) \\ \hline

      $\hat{\Delta}_{(\infty,\beta)}$ & $\Delta_1 [ 1 + O(1/\beta)] $ & $ \Delta_1 [ 1 + O(1/\beta^{\frac{1}{a+1}})] $ & $ \sim 1/\beta$ & $ \sim 1 / \beta^{ \frac{1}{a+1}} $ \\ 
        $\hat{\Delta}_{(n,\infty)}$ & $\Delta_1 [ 1 + O(1/n)] $ & $ \Delta_1 [ 1 + O(1/n^{\frac{1}{a+1}})] $ & 0 & $ \sim 1 / n^{ \frac{1}{a}}$ \\ \hline           
  \end{tabularx} }
\end{table}

  \vskip1.2em
  \underline{(ii) $ A_+ (\epsilon) \sim e^{- c / ( \epsilon - \Delta_1 )^a } \theta( \epsilon - \Delta_1 ) \quad (a,c>0)$}

  \vskip0.3em
Even when the spectral function vanishes exponentially at $\epsilon = \Delta_1 >0$, the gap estimator will be asymptotically unbiased. The convergence becomes slower than the power-law case:
\begin{align}
          \hat{\Delta}_{(\infty,\beta)} &\sim \sqrt{\frac{ U(a,c,2)}{ U(a,c,0) } } = \Delta_1 \left[ 1 + O \left( \frac{ 1 }{ \beta^{\frac{1}{a+1}} } \right) \right] , \\
    \hat{\Delta}_{(n,\infty)} &\sim \sqrt{\frac{ V(a,c,2n-1)}{ V(a,c,2n+1) } } = \Delta_1 \left[ 1 + O \left( \frac{ 1 }{ n^{\frac{1}{a+1}} } \right) \right] ,
  \end{align}
  where
  \begin{align}
            U(a,c,m) &\equiv \int_{\Delta_1}^{\epsilon_M} d\epsilon \, \frac{ e^{-
      \frac{c}{ ( \epsilon - \Delta_1 )^a } } \epsilon^{m} }{ \sinh \left( \frac{ \beta \epsilon }{ 2 } \right) } =:
    \int_{\Delta_1}^{\epsilon_M} d \epsilon \, e^{- u_m(\epsilon)}
    \approx \sqrt{\frac{2\pi}{ u''_m (\epsilon_m^*) } } e^{ - u_m(
      \epsilon_m^* ) } \sim \Delta_1^{-m} e^{ - m \, \beta^{ - \frac{ 1 }{ a + 1 } } } \quad ( \beta \gg 1 / \Delta_1 ) , \\
    V(a,c,m) &\equiv \int_{\Delta_1}^{\epsilon_M} d\epsilon \, e^{-
      \frac{c}{ ( \epsilon - \Delta_1 )^a } } \epsilon^{-m} =:
    \int_{\Delta_1}^{\epsilon_M} d \epsilon \, e^{- v_m(\epsilon)}
    \approx \sqrt{\frac{2\pi}{ v''_m (\epsilon_m^*) } } e^{ - v_m(
      \epsilon_m^* ) } \sim \Delta_1^{-m} m^{- \frac{ a+2 }{ 2( a + 1 ) }} e^{ - m^{ \frac{ a }{ a + 1 } } } \quad ( m \gg 1 ) ,
    \end{align}
  $u''_m(\epsilon)$ and $v''_m(\epsilon)$ are the second derivative, and the saddle-point approximation around its extremum $u'_m(\epsilon_m^*) = v'_m(\epsilon_m^*)=0$ were used, respectively.

  \vskip1.2em
  \underline{(iii) $ A_+ (\epsilon) \sim \epsilon^a \quad ( a > 0 )$}

  \vskip0.3em
  Let us consider the gapless case with a power-law form. From Eq.~(\ref{eqn:delta-conti}), the estimate will diverge since the gap is actually zero, but it takes a finite value for the case of finite $\beta$:
    \begin{align}
    \hat{\Delta}_{(n,\beta)} & \sim \sqrt{\frac{ W(a+3)}{ W(a+1) } } \sim 1/\beta \quad ( \beta \gg 1 ),
  \end{align}
  where
  \begin{align}
    W(m) &\equiv \int_{0}^{\epsilon_M} d\epsilon \, \frac{ \epsilon^m }{ \prod_{j=0}^n \left( \epsilon^2 + \omega_j^2 \right) } =: \int_0^{\epsilon_M} d\epsilon \, w_{m}(\epsilon) \sim w_{m}(\epsilon_{m}^* ) \sim \beta^{2(n+1)-m} \quad (\beta \gg 1) ,
  \end{align}
  because $w'_{m} (\epsilon_m^*) =0 \Rightarrow \epsilon_m^* \sim 1 / \beta$. Then $\hat{\Delta}_{(\infty,\beta)} \sim 1/\beta$ and $\hat{\Delta}_{(n,\infty)}=0$.

    \vskip1.2em
  \underline{(iv) $ A_+ (\epsilon) \sim e^{- c / \epsilon^a } \quad (a,c>0)$}

  \vskip0.3em
  At last let us consider the case where the spectrum is
  gapless and its function is exponential. The asymptotic form gains non-trivial exponents:
  \begin{align}
    \hat{\Delta}_{(\infty,\beta)} &\sim \sqrt{\frac{ X(a,c,2)}{ X(a,c,0) } } \sim \beta^{- \frac{1}{a+1}}, \\
        \hat{\Delta}_{(n,\infty)} &\sim \sqrt{\frac{ Y(a,c,2n-1)}{ Y(a,c,2n+1) } } \sim n^{- \frac{1}{a}} ,
  \end{align}
  where
  \begin{align}
    X(a,c,m) &\equiv \int_{0}^{\epsilon_M} d\epsilon \, \frac{ e^{-
      \frac{c}{ \epsilon^a } } \epsilon^{m} }{ \sinh \left( \frac{ \beta \epsilon }{ 2 } \right) } =: \int_{0}^{\epsilon_M} d \epsilon \, e^{- x_{m,p}(\epsilon)} \approx \sqrt{\frac{2\pi}{ x''_{m,p} (\epsilon_{m,p}^*) } } e^{ - x_{m,p}( \epsilon_{m,p}^* ) } 
    \sim e^{- \beta^{\frac{a}{a+1}}} \beta^{ - \frac{ a + 2 + 2m }{ 2(a+1)} } \quad (\beta \gg 1) \label{eqn:X} , \\
    Y(a,c,m) &\equiv \int_{0}^{\epsilon_M} d\epsilon \, e^{-
      \frac{c}{ \epsilon^a } } \epsilon^{-m} \approx \int_{0}^{\infty} d\epsilon \, e^{- \frac{c}{ \epsilon^a } } \epsilon^{-m} = \frac{c^{- \frac{m-1}{a}}}{a} \Gamma \left( \frac{ m - 1 }{ a } \right) \sim \left( \frac{m}{ace} \right)^{\frac{m}{a}} \quad (m \gg 1 ) \label{eqn:Y} .
    \end{align}
  In Eq.~(\ref{eqn:X}) $x''_m(\epsilon)$ is the second derivative and the saddle-point approximation around its extremum $x'_m(\epsilon_{m,p}^*)=0$ was used. In Eq.~(\ref{eqn:Y}), $\Gamma(t)$ is the gamma function.
  
\vskip1em
Remarkably, the asymptotic behavior of the gap
estimate is perfectly consistent also with the continuum spectrum in all of the
cases that we investigate here; that is
\begin{align}
  \lim_{\beta \rightarrow \infty} \lim_{n \rightarrow \infty} = \lim_{n \rightarrow \infty} \lim_{\beta \rightarrow \infty} \hat{\Delta}_{(n,\beta)} = \Delta_1 ,
\end{align}
including the gapless ($\Delta_1=0$) case.

The continuum spectrum that we have investigated appears in the
thermodynamic limit of many realistic quantum systems. The asymptotic
form should be considered to check the convergence of the gap
estimate and understand the spectrum structure. We note,
nevertheless, that a gapless mode in critical phases obeys the asymptotic expression obtained in the
discrete formalism. It is because the ratios between the first gap and
the higher gaps are kept even though the system size increases. In
other words, the low-energy-excitation spectrum is unchanged with the
scale transformation, which is certainly characteristic in critical
phases. We hence securely applied the discrete formalism to confirm
the convergence in the present analysis of the spin-Peierls model.

\begin{table}[b]
  \parbox{0.95\textwidth}{\caption{Mean $\bar{\mu}(T)$ and standard
      error $\bar{\sigma}(T)$ of the normalized quantity
      $\bar{\epsilon}=(\hat{\Delta}-\Delta_1)/\hat{\sigma}$ estimated
      from $2^{20} (\sim 10^6)$ Monte Carlo steps by the optimal fit
      and the gap estimators ($1\leq n \leq 5$) at temperature $T=1/6$
      and $1/8$. For the calculation of each value, 2048 independent
      simulations were run. The statistical errors indicated in the
      parenthesis were estimated by bootstrapping. \label{tab:error}}
  } \npdecimalsign{.}  \nprounddigits{3} \newcommand{\effddd}{1000}
  \newcommand{\effdd}{100}
      {\renewcommand{\arraystretch}{1.15}
  \begin{tabularx}{0.95\textwidth}{@{} *5{>{\centering\arraybackslash}X}@{}}
    \hline \hline
    Estimator   &  $\bar{\mu}(1/6)$  &  $\bar{\sigma}(1/6)$ & $\bar{\mu}(1/8)$ & $\bar{\sigma}(1/8)$ \\ \hline
    optimal fit
    & \nprounddigits{2}\numprint{-33.7457327819239}\MULTIPLY{0.477748308131856}{\effdd}{\sol}\nprounddigits{0}(\numprint{\sol})
    & \nprounddigits{2}\numprint{21.7587885197403}\MULTIPLY{0.328782303038717}{\effdd}{\sol}\nprounddigits{0}(\numprint{\sol})
    & \nprounddigits{2}\numprint{-2.22401388819175}\MULTIPLY{0.591463396759372}{\effdd}{\sol}\nprounddigits{0}(\numprint{\sol})
    & \nprounddigits{2}\numprint{18.9250832564716}\MULTIPLY{0.554020259515932}{\effdd}{\sol}\nprounddigits{0}(\numprint{\sol} \\
    \vskip-0.5em
    present approach\\
    
    $n=1$ & \numprint{4.06995535465651}\MULTIPLY{0.0224113718089131}{\effddd}{\sol}\nprounddigits{0}(\numprint{\sol})
    & \numprint{1.00794612319718}\MULTIPLY{0.0151771281912875}{\effddd}{\sol}\nprounddigits{0}(\numprint{\sol})
    & \numprint{2.46986291830128}\MULTIPLY{0.0223976575463904}{\effddd}{\sol}\nprounddigits{0}(\numprint{\sol})
    & \numprint{1.01166712378615}\MULTIPLY{0.0159134886784101}{\effddd}{\sol}\nprounddigits{0}(\numprint{\sol}) \\
    
    $n=2$
    & \numprint{0.489709847053302}\MULTIPLY{0.0220588013316961}{\effddd}{\sol}\nprounddigits{0}(\numprint{\sol})
    & \numprint{0.992077354971614}\MULTIPLY{0.0151285848931324}{\effddd}{\sol}\nprounddigits{0}(\numprint{\sol})
    & \numprint{0.168109270637345}\MULTIPLY{0.0223220577986323}{\effddd}{\sol}\nprounddigits{0}(\numprint{\sol})
    & \numprint{1.00483552883548}\MULTIPLY{0.0158808948717973}{\effddd}{\sol}\nprounddigits{0}(\numprint{\sol}) \\
    
    $n=3$
    & \numprint{0.130785144617151}\MULTIPLY{0.0220424861203509}{\effddd}{\sol}\nprounddigits{0}(\numprint{\sol})
    & \numprint{0.993267531803509}\MULTIPLY{0.0152654332537938}{\effddd}{\sol}\nprounddigits{0}(\numprint{\sol})
    & \numprint{0.0218682648454119}\MULTIPLY{0.0223197717603327}{\effddd}{\sol}\nprounddigits{0}(\numprint{\sol})
    & \numprint{1.00337683758954}\MULTIPLY{0.0161692252485325}{\effddd}{\sol}\nprounddigits{0}(\numprint{\sol}) \\
    
    $n=4$
    & \numprint{0.0569813926595412}\MULTIPLY{0.0219946927615905}{\effddd}{\sol}\nprounddigits{0}(\numprint{\sol})
    & \numprint{0.993059526123447}\MULTIPLY{.0153173852601041}{\effddd}{\sol}\nprounddigits{0}(\numprint{\sol})
    & \numprint{-0.00269828805465399}\MULTIPLY{0.0223090368531395}{\effddd}{\sol}\nprounddigits{0}(\numprint{\sol})
    & \numprint{1.00229638928027}\MULTIPLY{0.0164041862384116}{\effddd}{\sol}\nprounddigits{0}(\numprint{\sol}) \\
    
    $n=5$
    & \numprint{0.0334909578014074}\MULTIPLY{0.0219485753979585}{\effddd}{\sol}\nprounddigits{0}(\numprint{\sol})
    & \numprint{0.992539220141338}\MULTIPLY{0.0153153827228737}{\effddd}{\sol}\nprounddigits{0}(\numprint{\sol})
    & \numprint{-0.0113676146106986}\MULTIPLY{0.0223004052596155}{\effddd}{\sol}\nprounddigits{0}(\numprint{\sol})
    & \numprint{1.00181524457261}\MULTIPLY{0.0165451091344101}{\effddd}{\sol}\nprounddigits{0}(\numprint{\sol}) \\ \hline \hline
  \end{tabularx} }
\end{table}
\section{Comparison of error-bar estimations}
We will show the comparison of the error-bar estimations for the
relevant spin-Peierls model [Eq.~(\ref{eqn:sp-model}) for $L=4$,
  $\omega=4$, $\lambda=1/2$, $D$(cutoff)$=4$, in the main text]
between the gap estimators and the optimal fit. In the fitting method,
an optimal $\tau^*_1$ is selected so that the gap error is minimized
by the linear regression of $\log C(\tau)$ for $\tau_1 \leq \tau \leq
\beta/4$. To check the validity of the error bar, we show the mean and
the standard error of the normalized quantity
$\bar{\epsilon}=(\hat{\Delta}-\Delta_1)/\hat{\sigma}$ in
Table~\ref{tab:error}, where $\Delta_1 \simeq 1.111388$ is the exact
gap value, $\hat{\Delta}$ and $\hat{\sigma}$ are the gap and its error
bar, respectively, estimated from each simulation of $2^{20}(\sim
10^6)$ Monte Carlo steps. Then the mean and the standard error of the
normalized quantity were calculated from 2048 independent
simulations. The mean should approach zero if the estimation is
unbiased, and the standard error should become one if the error bar is
appropriately estimated.  While the bias of the optimal fit becomes
smaller as temperature decreases, the standard error is significantly
large as shown in the table. It is because the correlation between
data at different imaginary times is ignored and the estimated error
bar is improperly too small. This inappropriate error-bar estimation
causes the deviation from the exact value to become typically
$20\sigma$ ($\sim2.22+18.93$, which are in the table) even at
$T=1/8$. If the error bar is naively used for another analyses, e.g.,
the extrapolation to the thermodynamic limit as we showed in the main
text, it might end up a wrong conclusion.  On the other hand, our
higher-order gap estimator is asymptotically unbiased, and the error
bar is reliably estimated (see Table~\ref{tab:error}). This is a clear
advantage of the present approach, and it makes the precise analysis
of the criticality possible as demonstrated in the present paper.

We note that more careful statistical analyses like
bootstrapping\,\cite{DavisonH1997} would improve the estimate of the error bar
even in the fitting method (also in the optimal fit). It
is, however, necessary to run an additional (Monte Carlo) simulation that
is somewhat costly for users. Moreover, one has to make sure that the
number of (almost) independent bins is large enough. Otherwise the
variance of the error bar becomes improperly large and the result of
the bootstrapping is unreliable. The needed number of floating-point
operations scales as $O((nm + \alpha m)M)$ once a fitting
region~($\tau$) is fixed, where $n$ ($\sim 10^2$--$10^3$ typically) is
the number of bins, $m$ ($\sim 10^2$--$10^3$) is the number of data
points (at different $\tau$), $\alpha m$ is the computational cost for
a regression, $\alpha$ ($\sim 10$--$10^3$) depends on regression
schemes and the number of parameters, and $M$ ($\sim 10^3$--$10^4$) is
the number of bootstrap samples. Then the whole process including the
fitting-region~($\tau_1$) optimization costs $O((nm+ \alpha m^2)M)$.
It would take a few hours for large system sizes with large number of
data since the needed data points ($\tau$) will be proportional to the
system length in critical phases. On the other hand, it is much
easier, in our approach, to calculate a valid error bar simply by the first-order jackknife method\,\cite{S-Berg2004}
that costs only $O(n)$. Our approach is, thus, more accurate than
simple (or naive) fitting methods like the optimal fit as we showed,
and also handier and more straightforward than the bootstrapping.

\section{Feasibility of higher-order gap-estimation method}
We will discuss the variance of the gap
estimators~(\ref{eqn:delta-n}) and the feasibility of our
approach. Because the distribution of an average of Monte Carlo samples
will be Gaussian according to the central limit theorem, the
estimator $\hat{\Delta}_{(n,\beta)}^2$ takes the form of the ratio
between the two Gaussian distributions. The variance of the ratio estimator
$R=X/Y$, where $X \sim N(\mu_x, \sigma_x^2)$ and $Y \sim N(\mu_y,
\sigma_y^2)$, is expressed\,\cite{DavisonH1997} as ${\rm Var}(R)
\approx \sigma_y^2 \mu_x^2 / \mu_y^4 + \sigma_x^2 / \mu_y^2 - 2 \rho
\sigma_x \sigma_y \mu_x / \mu_y^3$, where $\rho$ is the correlation
coefficient between $X$ and $Y$. Then the statistical error the gap
estimate becomes
\begin{align}
\sigma_{\hat{\Delta}_{(n,\beta)}} \equiv \sqrt{ {\rm
    Var}(\hat{\Delta}_{(n,\beta)}) } \sim \sqrt{ {\rm
    Var}(\hat{\Delta}_{(n,\beta)}^2) } \sim ( \mu_D^{-1} + an ) / \sqrt{M}
\sim ( z^{2n} + an ) /\sqrt{M} \label{eqn:var} ,
\end{align}
where $\mu_D = \langle \sum_{k=0}^n x_{n,k} \tilde{C}(\omega_k)
\rangle$, $z \equiv \Delta_1 / \omega_1 \equiv (\beta \Delta_1 / 2
\pi)$, $M$ is the number of Monte Carlo steps, and $a$ is a positive
real constant. The symbol $\langle A \rangle$ means the statistical
average of $A$. The term of coefficient $a$ comes from the fact that
$\hat{\Delta}_{(n,\beta)}$ needs the Fourier components
at $\omega_j$ $(0 \leq j \leq n)$.  A typical value of $a$ in the
present paper can be estimated around $0.01$--$0.1$. Thus,
from~Eq.~(\ref{eqn:var}), the statistical error of the gap estimate
does not increase much as $n$ increases for $z<1$. On the other hand,
it rapidly grows as $n$ for $z>1$. Therefore the higher gap estimators work only for $z \lesssim O(1)$.

Next let us estimate the value of needed $z$ (or $\beta$) so that
$\hat{\Delta}_{(\infty,\beta)} / \Delta_1$ is close enough to 1. For
$\beta > 1 / \Delta_1$, we need to take into account, in
Eq.~(\ref{eqn:delta-inf}), only the terms with $\ell' = 0$. Then
$\beta \Delta_{\rm ex} / 2 =\beta ( \Delta_{\ell} - \Delta_1 ) / 2
\equiv \beta \Delta_1 ( r_{\ell} - 1 ) / 2 = \pi z ( r_{\ell} - 1 )$,
where $r_{\ell}\equiv \Delta_{\ell} / \Delta_1$. Suppose we want to
achieve the order of the systematic error as $\exp( - \pi z ( r - 1 )
) \lesssim 10^{-m}$. Equivalently, $z \gtrsim m \ln 10 / \pi ( r - 1
)$. In the case where $m=r=5$, $z \gtrsim 5 \ln 10 / 4 \pi \approx
0.916$, which is feasible to use the higher-order estimators
from the above error argument. For the $n$ convergence, according to
Eq.~(\ref{eqn:finite-n}), the needed order is $n \gtrsim n_{m} \sim
\alpha z^2$, where $\alpha$ ($\sim$$5$) is a constant. Then, now, $n_m
\sim \alpha z^2 \sim 5$, which is also feasible to calculate.

Actually, this example is the case for the present transition point of
the spin-Peierls model. The excitation energy is expressed as
$\Delta_{\ell} = 2 \pi v x_{\ell} / L = 2 \pi v ( x_1 + 2 ( \ell - 1 )
) / L$ at the (1+1)-dimensional critical systems with conformal
invariance, where $x_{\ell}$ is the scaling dimension, because the
higher excitation will come from the descendant field of the primary
field corresponding to the first excitation (with the same wave
number). Therefore, the ratio of the gaps becomes $r_{\ell} \equiv
\Delta_{\ell} / \Delta_1 = ( x_1 + 2 ( \ell - 1 ) ) / x_1$. Then, as
the smallest $r$, $r_2=5$ in the present study since $x_1=1/2$. In
general, for (1+1)-dimensional conformal systems, the scaling
dimension of a relevant field $x_1 < 2$ and $r_2 = ( x_1 + 2 ) / x_1 >
2$, which indicates needed $z \sim O(1)$. Thus the present gap analysis is
expected to work generally for the analysis of conformal invariant
phases and transition points.

To sum up, we propose a recipe for the precise gap estimation in
general. (i) First, roughly estimate the gap as $\Delta_1 \approx
\tilde{\Delta}_1$, say with 10\% accuracy, by the second moment
estimator~(\ref{eqn:delta-1}) or some way at low enough temperature
$\beta \gtrsim 2 \pi / \Delta_1$. The consistency can be checked by $
\beta \gtrsim 2 \pi / \tilde{\Delta}_1$. (ii) Set temperature
$\beta=\tilde{\beta}^* \equiv 2 \pi / \tilde{\Delta}_1$, equivalently
$z \approx 1$. (iii) Calculate the higher-order gap estimate
$\hat{\Delta}_{(n,\tilde{\beta}^*)}$ with $n \sim 5$ as the final gap
estimation. Then the (relative) systematic and statistical error can
become as small as $10^{-5}$--$10^{-4}$. Note that the actual
statistical error is $O(1/\sqrt{M})$, where $M$ is the number of Monte
Carlo steps, so $M$ naturally needs to be $10^8$--$10^{10}$ to achieve
the precision. The important points here are the systematic error is
securely smaller than an achievable statistical error and the total of
the systematic and statistical errors will be actually reduced to such
a small number. This recipe indeed works for the test case in the main
text, where the higher-order gap estimate converges well at $T=1/6$,
equivalently $z=6\Delta_1/2\pi=1.0613... \approx 1$, as shown in
Fig.~\ref{fig:tg}. We followed the recipe for the analysis of the
spin-Peierls model in the present study.

\end{document}